\documentclass[a4paper]{article}

\usepackage{INTERSPEECH2019}

\newcommand\blfootnote[1]{%
  \begingroup
  \renewcommand\thefootnote{}\footnote{#1}%
  \addtocounter{footnote}{-1}%
  \endgroup
}

\usepackage{graphicx}
\usepackage{subcaption}

\title{Voicy: Zero-Shot Non-Parallel Voice Conversion in Noisy Reverberant Environments}
\name{Alejandro Mottini*, Jaime Lorenzo-Trueba*, Sri Vishnu Kumar Karlapati, Thomas Drugman}
\address{
 Amazon.com}
\email{(amottini,truebaj,srikarla,drugman)@amazon.com}

\begin{document}

\maketitle
\begin{abstract}
Voice Conversion (VC) is a technique that aims to transform the non-linguistic information of a source utterance to change the perceived identity of the speaker. While there is a rich literature on VC, most proposed methods are trained and evaluated on clean speech recordings.  However, many acoustic environments are noisy and reverberant, severely restricting the applicability of popular VC methods to such scenarios. To address this limitation, we propose Voicy, a new VC framework particularly tailored for noisy speech. Our method, which is inspired by the de-noising auto-encoders framework, is comprised of four encoders (speaker, content, phonetic and acoustic-ASR) and one decoder. Importantly, Voicy is capable of performing non-parallel zero-shot VC, an important requirement for any VC system that needs to work on speakers not seen during training. We have validated our approach using a noisy reverberant version of the LibriSpeech dataset. Experimental results show that Voicy outperforms other tested VC techniques in terms of naturalness and target speaker similarity in noisy reverberant environments.
\end{abstract}

\blfootnote{* Equal contribution}

\section{Introduction}
\label{sec:intro}

Voice Conversion (VC) is the task of modifying an utterance from a source speaker to make it sound like it was uttered by a target speaker, while preserving the original linguistic content  \cite{lorenzo2018voice}. VC is a key component of many modern applications, including text-to-speech (TTS) \cite{cotescu2019voice}, speech enhancement \cite{toda2012statistical}, and speaking assistance \cite{nakamura2012speaking} systems. Due to its success in these fields, VC has been studied extensively in recent years \cite{mohammadi2017overview}. 

However, despite their success in generating realistic  samples, most current VC approaches have two important limitations. First, they are trained and evaluated on clean speech recordings, such as LibriSpeech  \cite{panayotov2015librispeech} or VCTK \cite{vctk}. This is a shortcoming, since most real acoustic environments are noisy and reverberant. Second, not all methods can perform non-parallel zero-shot conversion \cite{qian2019zero, kameoka2018stargan}. These shortcomings limit their applicability to certain industrial use-cases, such as applying VC to noisy utterances captured by voice-controlled virtual assistants. In such scenarios, a production VC system should work well in more realistic acoustic conditions (noisy and reverberant) and be robust to microphones with different characteristics. In addition,  VC methods should be capable of transforming utterances from speakers not seen during training, and be able to change the speaker’s identity while preserving the quality and naturalness of the speech. Finally, a production VC system should be scalable and robust.

In this paper we propose Voicy, a new VC method that fulfils the desired characteristics outlined before. Our approach, based on de-noising auto-encoders \cite{vincent2008extracting} and the AutoVC model \cite{qian2019zero}, is especially tailored for noisy reverberant speech (both as source and target), and capable of performing non-parallel zero-shot VC. Importantly, the proposed phonetic and acoustic-ASR encoders, an improvement over the original AutoVC formulation, significantly improves the intelligibility of the converted speech in noisy conditions. Moreover, since Voicy is based on auto-encoders, it is more robust and easier to train than other GAN- \cite{kameoka2018stargan} and  Flow-based \cite{serra2019blow} approaches. 

To validate our approach, we have created a noisy reverberant version of the LibriSpeech dataset \cite{panayotov2015librispeech}, and used it to train and test both our method and other selected baselines. Results show that Voicy outperforms other VC techniques in terms of naturalness and target speaker similarity in noisy reverberant environments. Converted speech samples are provided here \footnote{https://github.com/alexa/amazon-voice-conversion-voicy}.


\section{Related Work}
\label{sec:relWork}

Based on the required training data, VC methods can first be characterized as parallel or non-parallel. Models in the first category \cite{abe1990voice,takashima2012exemplar} depend on a training set of aligned speech pairs of source and target speakers uttering the same phrase. Conversely, non-parallel models only require source and target speaker's utterances, but they do not need to be aligned or match in terms of content. Non-parallel VC techniques can further be characterized as either zero-shot or not, depending on their ability to transform utterances of speakers unseen during training. Naturally, non-parallel zero-shot VC is the most challenging but valuable framework, and is therefore the focus of this work.

Regardless of the type of required data, the actual conversion technique behind each VC method varies. Some methods rely on traditional statistical approaches such as Gaussian Mixture models (GMM) \cite{stylianou1998continuous}, while newer techniques use Deep-Learning-based approaches. Within this family, different approaches exist, most notably, Generative-Adversarial-Network-based (GAN) \cite{kameoka2018stargan,fang2018high}, Variational-Auto-Encoder-based (VAE) \cite{kameoka2018acvae}, Auto-Encoder-based \cite{qian2019zero}, and Flows-based \cite{serra2019blow} models. Each family has its own trade-off between conversion quality, complexity and ease of training. In general, Auto-Encoder-based models appear to have the best trade-off \cite{qian2019zero}. In particular, \cite{saito2018non} proposes a variational-autoencoder method conditioned on the phonetic contents of utterances, but it is not zero-shot, and was only evaluated on a small dataset of clean utterances. 


Finally, a small body of work is dedicated to VC in noisy environments. In \cite{takashima2013noise}, a parallel exemplar-based VC model is proposed. Another approach is \cite{miao2020noise}, where a speech-enhancement-based technique that applies two different filtering methods to suppress noise is proposed. Then, a traditional BLSTM-VC model \cite{sun2015voice} is used to convert the filtered utterances. Although successful, these approaches have several shortcomings, including their inability to perform zero-shot conversion, and the fact that the presented results are compared against relatively weak baselines (GMM-based VC).

\section{Voicy: Our Proposed Method}


\subsection{General System Description}


Voicy is comprised of 5 modules (see Figure \ref{fig:ASR-embedding-train-inference}), and uses two representations of an utterance: Mel-spectrogram and transcription, represented using phonemes. 

Our speaker encoder follows \cite{wan2018generalized}. We use a model pre-trained on VCTK \cite{vctk}, which remains fixed during training. In addition, our content encoder is inspired by AutoVC \cite{qian2019zero}, but with minor modifications to the hyper-parameters.

The phonetic encoder, with its architecture detailed in Figure \ref{fig:ASR-embedding-train-inference}, is responsible for encoding the sequence of phonemes into a sentence-level representation of the text. As such, the content encoder is not forced to be in charge of both maintaining the linguistic information and the prosody of the speech. The main effect of adding this component is having a specific module in charge of intelligibility, which improves the naturalness of the converted speech (see Section \ref{ssec:Results}). 

Finally, the ASR module,  comprised of CNN and bi-LSTM layers (Figure \ref{fig:ASR-embedding-train-inference}), learns to predict the phonetic embeddings produced by the phonetic encoder, but working in the audio instead of the text domain. As such, once our model is trained, the phonetic encoder can be substituted by the ASR module, removing the need for the textual representation of the input utterance during inference. This makes the applicability of this approach in production more viable. As an alternative, one could also consider a system were a standard ASR model that automatically generates a textual representation of the utterances, which could then be encoded thanks to the phonetic encoder. However, to maximize performance, we have opted for our design that uses the best of both worlds: use transcriptions if available, or an ASR module when not. This motivated our choice of having 2 modules instead of just one.



The decoder's architecture follows \cite{qian2019zero}, and is comprised of GRU and CNN layers. It receives the output of the speaker, content, and phonetic or ASR encoders (depending on the stage), and outputs the converted spectrogram. We can interpret its 3 inputs as: (1) what is being uttered (linguistic information captured by the phonetic/ASR embedding), (2) who uttered it (speaker identity captured by the speaker embedding), (3) how it is uttered (prosody information captured by the content encoder). Finally, the Universal WaveRNN-like Vocoder  \cite{lorenzo2018towards} is used to convert the produced Mel-spectrogram into a  waveform. 


\subsection{Architecture Description}


Let us first define the tuple $(S,Z,A)$ representing speaker $S$, content (phonetic and prosodic) $Z$, and  audio segment $A$. Let us now take two such tuples, $(S_1,Z_1^i,A_1^i)$ and $(S_2,Z_2^k,A_2^k)$, where the first corresponds to the $i$-th tuple of the source speaker 1, and the second to the $k$-th tuple of the target speaker 2. The goal of any VC system is to produce the output utterance $\hat{A}_{1\rightarrow 2}^i$ that keeps the content of $A_1^i$, while changing the perceived speaker to $S_2$. Since we tackle the zero-shot VC problem, $S_1$ or $S_2$ do not need to be part of the training set.  

To achieve this, Voicy uses five modules: speaker encoder $E_s(.)$,  content encoder $E_c(.)$, phonetic encoder $E_{ph}(.)$,  acoustic-ASR encoder $E_{ASR}(.)$, and  decoder $D(. , .)$.  Both the phonetic and acoustic-ASR encoders are improvements over AutoVC \cite{qian2019zero}, which, along with the use of the de-noising auto-encoder technique, allow our model to perform VC in a noise-robust manner. 

\begin{figure}[!h]
    \centering
    \begin{subfigure}[t]{0.5\textwidth}
        \centering
   \includegraphics[scale=0.55]{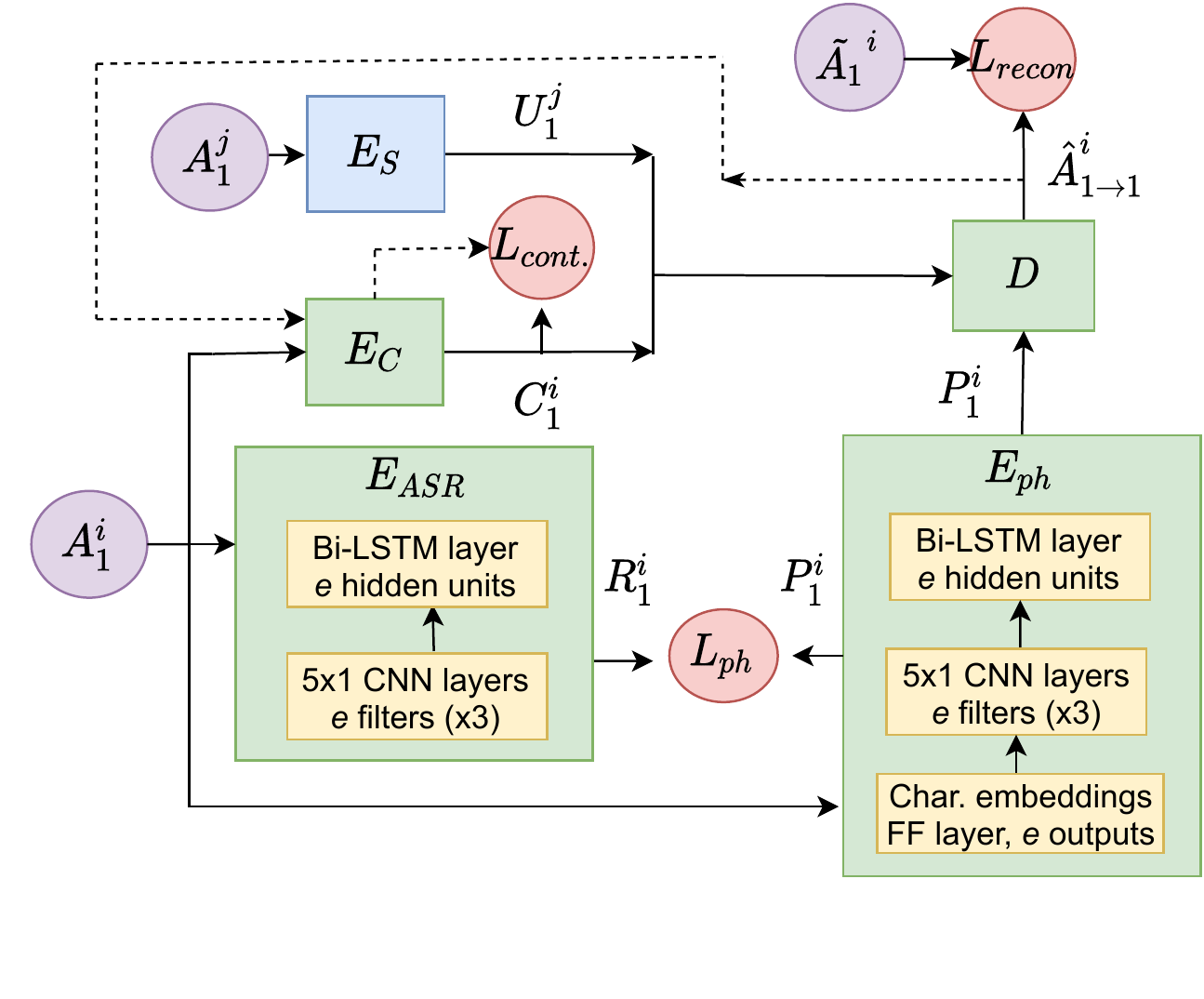}
  \includegraphics[scale=0.55]{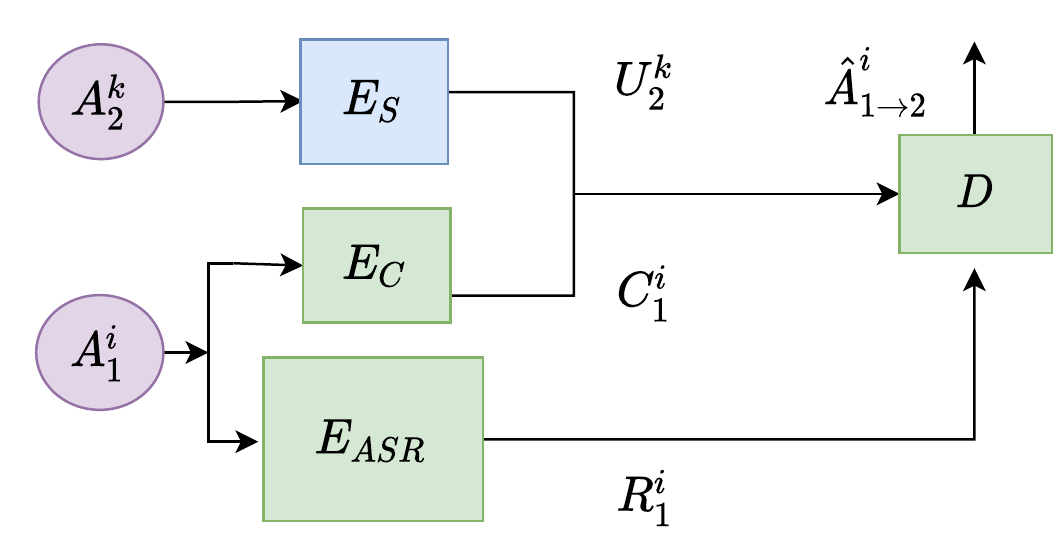}
    \end{subfigure}%
 \caption{Voicy during training (top) and test (bottom) phase. For simplicity, $f_{ML}(A)$ and $f_{ph}(A)$ are not shown. The speaker encoder $E_S$ (in blue) is pre-trained and fixed during training.}
  \label{fig:ASR-embedding-train-inference}
\end{figure}

More concretely, given an utterance $A$, we represent it using two modalities: its Mel-spectrogram $ML$, and its transcription, represented using phonemes $PH$. For simplicity, we denote $ML=f_{ML}(A)$ and $PH=f_{ph}(A)$. Then, let us represent the input/output of each module as:
\begin{equation}
\begin{split}
C = E_c(ML), U = E_s(ML) \\
 R = E_{ASR}(ML), P = E_{ph}(PH) \\
 \hat{A}_{. \rightarrow  .} = D(C,U,R,P)\\
\end{split}
\end{equation}

\noindent where D receives either R or P, but not both. During training (Figure~\ref{fig:ASR-embedding-train-inference}), we use inputs  $(S_1,Z_1^i,A_1^i)$ and  $(S_1,Z_1^j,A_1^j)$, representing two different utterances from the same speaker $S_1$. Following the de-noising auto-encoder methodology, and leveraging the parallel corpus of clean and noisy data we constructed (see Section \ref{sec:val}), we also consider tuple $(S_1,Z_1^i,\tilde{A_1^i})$, where $\tilde{A_1^i}$ is a clean version of $A_1^i$, always available during training. Then, for each training utterance $A_1^i$ containing either clean, noisy, or noise reverberant speech, the model produces $C_1^i = E_c(f_{ML}(A_1^i))$, $U_1^j= E_s(f_{ML}(A_1^j))$, $R_1^i = E_{ASR}(f_{ML}(A_1^i))$, $P_1^i  = E_{ph}(f_{PH}(A_1^i))$ and $\hat{A}_{1 \rightarrow  1}^i = D(C_1^i,U_1^j,P_1^i)$. As such, the speaker and liguistic embeddings are extracted from both clean and noisy data. We then compute the total loss:


\begin{equation}
\begin{split}
L = L_{recon} + \beta L_{phonetic} + \lambda L_{content}  \\
L_{recon} = \lVert \hat{A}_{1 \rightarrow  1}^i - \tilde{A_1}^i  \rVert_2 \\
L_{phonetic} = \lVert R_1^i - P_1^i  \rVert_1 \\
L_{content} =   \lVert E_c(\hat{A}_{1 \rightarrow  1}^i ) - C_1^i  \rVert_1  \\
\end{split}
\end{equation}

\noindent with $\lambda$ and $\beta$ hyper-parameters of the model. $L_{recon}$, $L_{phonetic}$ and $L_{content}$ are loss functions computed using the inputs and outputs of the difference modules, and are presented in Figure \ref{fig:ASR-embedding-train-inference}. Speaker encoder $E_s(.)$ is assumed to be pre-trained and remains fixed during training. It is used to handle both clean and noisy reverberant speech. The model only sees utterances from a single speaker (when working in batches, many speakers are used) during training, and the reconstruction loss is always computed between the output of the decoder and the clean version of the input. This teaches the encoders to be  robust to noise.

Once trained, the model can convert utterances from source to target speakers. For this step (see Figure~\ref{fig:ASR-embedding-train-inference}), let us again consider tuples of two speakers $(S_1,Z_1^i,A_1^i)$ and $(S_2,Z_2^k,A_2^k)$. We then use the trained modules to compute: $C_1^i = E_c(f_{ML}(A_1^i))$, $U_2^k= E_s(f_{ML}(A_2^k))$, $R_1^i = E_{ASR}(f_{ML}(A_1^i))$ and $\hat{X}_{1 \rightarrow  2}^i = D(C_1^i,U_2^k,R_1^i)$.  As we can see, the phonetic encoder $E_{ph}$ is no longer needed (no transcription needed), and $E_{ASR}$ takes its place since it learned how to approximate its behavior. As such,  we can convert utterances for which we do not have the transcription. Moreover, the speaker encoder $E_s$ now receives an utterance from the target speaker $S_2$, and its output is fed to the decoder to reconstruct the speech as if it was uttered by $S_2$.




\section{Experimental Validation}
\label{sec:val}

\subsection{Experimental Protocol}

Using LibriSpeech as the basis, we have first created reverberant utterances using the Aachen Impulse Response (AIR) Database \cite{jeub2009binaural}, which contains over 200 room impulse responses for diverse settings. Due to the sizes and properties of the rooms, the reverb can be significant, with reverberation time (T60) ranging from 0.12 to 1.25 seconds. For this set of transformations, no noise is added.  In addition, we have created a second set of utterances containing both noise and reverberation using Pyroomacoustics \cite{scheibler2018pyroomacoustics}, a room acoustics simulation package. Using this simulator, we created 3D shoebox rooms of different sizes, trying to approximate the dimensions of realistic home rooms. Then, for each room configuration, we have added 3 audio sources: (1) clean LibriSpeech utterance, (2) white noise of different levels, and (3) external noises selected at random from the DEMAND dataset \cite{thiemann2013diverse}. This dataset is comprised of real-world noise from a variety of settings. The position of these 3 sources, along with the position of the (virtual) microphone, are varied randomly for each room/utterance. The resulting utterances have on average less reverberation than the previous set (due to the room size), but have two noise sources. 

Both approaches are applied to LibriSpeech train-clean-100 (250 speakers, 100 hours), train-clean-360 (920 speakers, 360 hours) and dev-clean (40 speakers, 5.3 hours). Both train-clean-100 and train-clean-360, along with their reverberant and noisy reverberant versions, are combined into a single training set. To create the  evaluation set (see Section \ref{ssec:Results}), we first combine the dev-clean and its reverberant and noisy-reverberant versions. We then randomly select a subset of 400 utterances, covering all speakers and noise levels (clean, reverb, and noisy-reverb). The SNR distributions in both training and test sets match, with a maximum SRN of 35 dB (for the clean Librispeach audios), a minimum of  -2.2 dB, and an average of 16 dB. All audio has a sampling frequency of 24kHz. Finally, we chose 2 random target speakers from the training set, one female and one male, and used each  VC model to transform the 400 utterances into them. 

Mel-spectrograms are extracted using the LibRosa library \cite{mcfee2015librosa}, with 80 coefficients and frequencies ranging from 50 Hz to 12 kHz. To obtain the phonetic representation, we have used the transcription provided with LibriSpeech, and the Montreal Forced Aligner \cite{mcauliffe2017montreal}.

As baselines, we have considered four methods besides Voicy (referred to as Proposed in Section \ref{ssec:Results}). First,  AutoVC \cite{qian2019zero}, which inspired our work. In addition, to compare our model against a speech-enhancement-based approach, we trained AutoVC on the original clean LibriSpeech, and used it to transform de-noised de-reverbed test utterances obtained by applying the LogMMSE \cite{ephraim1985speech} and the Weighted Prediction Error \cite{nakatani2010speech} speech enhancement methods to our noisy evaluation set. This methodology is referred to as Preproc. Moreover, we considered StarGAN-VC \cite{kameoka2018stargan}, an established GAN-based VC model. StarGAN-VC uses a one-hot representation of the speakers, and needs to see utterances from both training and test set speakers during training. To address this problem, we have taken 10\% of the test set and added it to StarGAN-VC training set. This 10\% is not included in the evaluation set of the methods. Finally, we perform a simple ablation study by removing the acoustic-ASR encoder from our architecture and using this model as a baseline. We refer to this approach as Phonetic. Since this variant uses the phonetic encoding during inference, and is fed the ground-truth transcription, its performance should provide an upper bound of the metrics.  We have used Wilcoxon signed-rank test to do pairwise comparisons between the different approaches.  Hyper-parameters were optimized using random search to maximize the perceived quality of samples under informal listening. 

To quantify the performance of the methods, we run two perceptual evaluations inspired by past voice conversion challenges~\cite{lorenzo2018voice}, looking at Naturalness and speaker Similarity. The evaluations were   crowd-sourced on Amazon Mechanical Turk, and designed according to MUltiple Stimuli with Hidden Reference and Anchor (MUSHRA)~\cite{recommendation2001method}, but without forcing any system to be rated as 100. Each evaluator rated 20 screens, and selected the Naturalness and Similarity to the target speaker using a 0 to 100 scale. In each evaluation screen, listeners were presented with samples from the 5 systems, and with recordings of the target speaker as hidden reference. A different random recording of the target speaker was provided as explicit reference. We collected 3 scores per utterance, for both target speakers.


\subsection{Results}
\label{ssec:Results}

We first evaluated the methods in clean acoustic conditions (original LibriSpeech data, no noise nor reverberation). Results (not presented for brevity) showed that the performance difference between the methods is minor for both metrics, with Phonetic outperforming the others due to its ability to leverage the phonetic information available during inference. In this scenario, Voicy does not outperform the other methods.

\begin{figure}[!h]
    \centering
    \begin{subfigure}[t]{0.5\textwidth}
        \centering
  \includegraphics[scale=0.18]{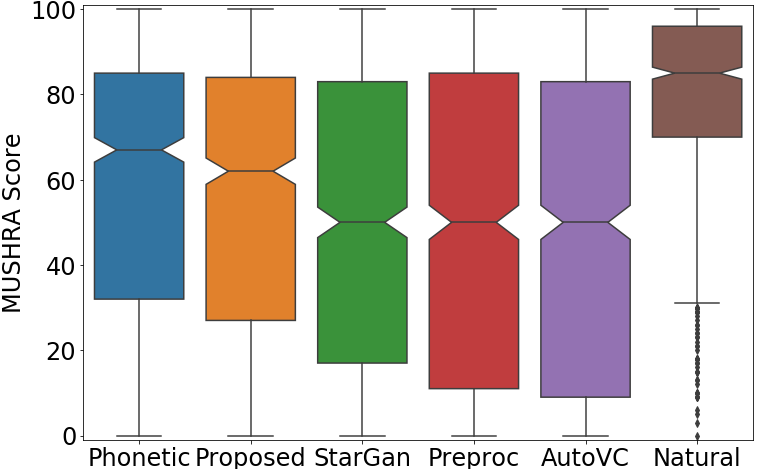}
   \includegraphics[scale=0.18]{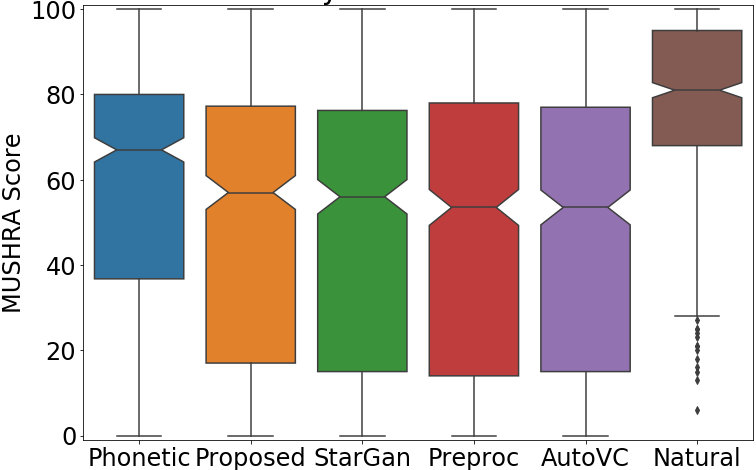}
    \end{subfigure}%
 \caption{Naturalness (top) and Similarity (bottom) of the systems for reverberant utterances.}
      \label{fig:reverb_mushra}
\end{figure}
\begin{figure}[!h]
    \centering
    \begin{subfigure}[t]{0.5\textwidth}
        \centering
   \includegraphics[scale=0.18]{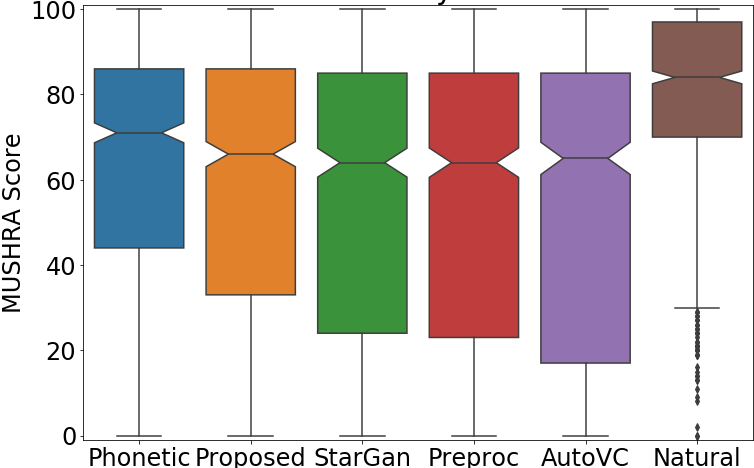}
   \includegraphics[scale=0.18]{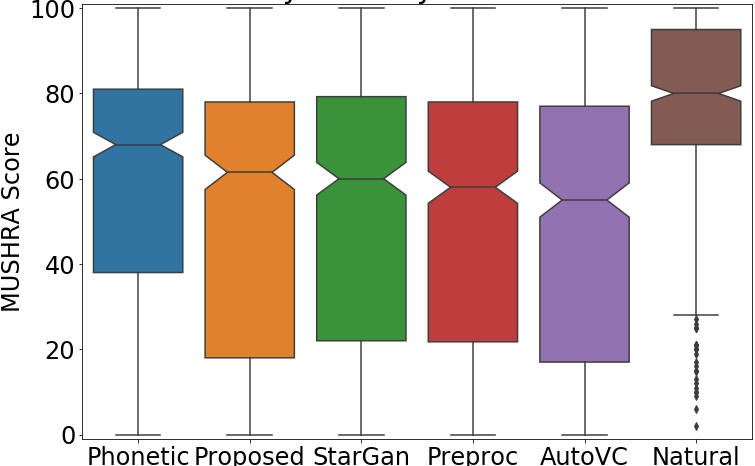}
    \end{subfigure}%
 \caption{Naturalness (top) and Similarity (bottom) of the systems for noisy reverberant  utterances.}
      \label{fig:noisyreverb_mushra}
\end{figure}

However, when analyzing the results for the reverberant (Figure \ref{fig:reverb_mushra}) and noisy reverberant utterances (Figure \ref{fig:noisyreverb_mushra}), we observe a clear difference between our systems and the baselines. Once again, Phonetic outperforms all others for both metrics, while the Proposed approach significatively outperforms the remaining baselines in terms of Naturalness on both reverberant and noisy reverberant conditions, likely due to the added information provided by the acoustic encoder during inference. Results are more mitigated for Similarity due to the high variance of the results for all methods. The relatively  large variances in performance is source/target dependent, but also on the level and type of noise. Models that handle noise better have significant less variance. Moreover, the use of Mechanical Turkers can add significant variability to the scores, even if precise instructions are given to them. Nevertheless, statistical analyses of the pair-wise comparisons between the systems show that the only non-significant improvement between Voicy and the baselines is for Similarity for reverberant conditions. For example, when comparing Proposed and Star-GAN, we observe $p=1.35E-5$ and  $p=8.46E-6$ for Naturalness in reverberant and noisy-reverberant conditions respectively, and $p=0.21$ and $p=0.038$ for Similarity. This could be due to the fact that the added phonetic context does not add any speaker disentangling information. In addition, we observe how AutoVC's performance can improve by applying speech-enhancement techniques as pre-processing (Preproc), particularly for the noisy reverberant utterances.  The reader is encouraged to listen to converted samples in different noise conditions (see Section \ref{sec:intro}).

\begin{figure}[!h]
    \centering
    \begin{subfigure}[t]{0.5\textwidth}
        \centering
\includegraphics[scale=0.40]{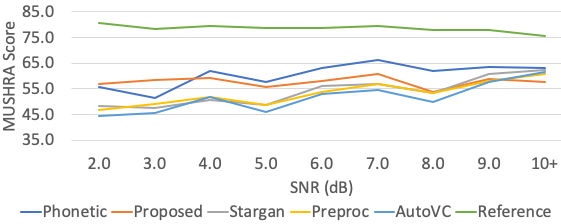}
    \end{subfigure}%
    \\
    \begin{subfigure}[t]{0.5\textwidth}
        \centering
\includegraphics[scale=0.40]{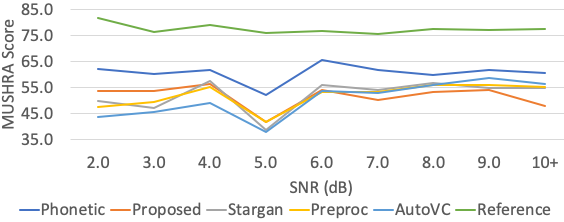}
    \end{subfigure}
    \caption{Average score for Naturalness (top) and Similarity (bottom) by SNR.}
      \label{fig:noisyreverb_snr}
\end{figure}

Moreover, we have estimated the SNR of the utterances using the Pysepm toolbox  \footnote{https://github.com/schmiph2/pysepm}, and analyzed the performance of the models for different noise levels. Results are presented in Figure \ref{fig:noisyreverb_snr}) for a SNR  range of interest (noisy conditions). Results show that our method outperforms the others (except Phonetic) in terms of Naturalness until 8 dB, and until 5 dB for Similarity. For higher levels,  Star-GAN and AutoVC appear to match or slightly outperform Voicy, which could be due to the fact the added phonetic encoder does not contribute to the speech reconstruction in  clean acoustic conditions. We also observe a degradation in performance for all of the systems at 5 dB SNR. We believe that during the random process of adding noise, more samples fell into the 5db noise bucket. The larger number of speakers and samples might have affected the performance of all models equally.


Overall, results concur in showing that Voicy outperforms the baselines in terms of Naturalness and Similarity in noisy reverberant environments, except for Phonetic, which has an "unfair" advantage during inference. The Proposed approach, with its acoustic encoder, tries to match its performance without the need for the phonetic representation during inference, but is unable to reach this upper bound. Nonetheless, it outperforms the other baselines. We believe these results could be improved using a  better pre-trained ASR module.

\section{Conclusions}

In this work we presented Voicy, a new VC model designed for use-cases that require noise/reverb robustness as well as zero-shot transformation, which is not possible with current VC approaches. Our architecture is comprised of five modules, including a phonetic and an acoustic-ASR encoder, which help improve the intelligibility of the converted speech in noisy environments. We have created a noisy reverberant version of the LibriSpeech dataset, and used to train and test both our method and four other baselines. Results show that Voicy outperforms the baselines in terms of naturalness and speaker similarity in noisy reverberant environments. In the future, we will improve our acoustic-ASR encoder, and study if there is prosodic leakage in the phonetic embedding.

\bibliographystyle{IEEEtran}

\bibliography{mybib}


\end{document}